\documentclass[10pt,journal]{IEEEtran}

\usepackage{cite}
\usepackage{graphicx}
\usepackage{algorithm}
\usepackage{algorithmic}
\usepackage{url}
\usepackage{amsmath}

\renewcommand{\algorithmicrequire}{\textbf{Initialization:}}
\renewcommand{\algorithmicensure}{\textbf{Iteration:}}

\begin{document}

\title{Design of a visible-light-communication enhanced WiFi system}

\author{Sihua Shao, Abdallah Khreishah, Moussa Ayyash, Michael B. Rahaim, Hany Elgala, \\Volker Jungnickel, Dominic Schulz, Thomas D.C. Little
\thanks{Sihua Shao and Abdallah Khreishah are with the Department of Electrical and Computer Engineering,
New Jersey Institute of Technology, email: ss2536@njit.edu, abdallah@njit.edu}
\thanks{Moussa Ayyash is with the Department of Information Studies, Chicago State University,
email: mayyash@csu.edu}
\thanks{Michael B. Rahaim, Hany Elgala, Thomas D.C. Little are with the Department of Electrical and Computer Engineering,
Boston University, email: mrahaim@bu.edu, helgala@bu.edu, tdcl@bu.edu}
\thanks{Volker Jungnickel and Dominic Schulzi are with Fraunhofer-HHI, email: volker.jungnickel@hhi.fraunhofer.de, dominic.schulz@hhi.fraunhofer.de}}

\maketitle
\begin{abstract}
Visible light communication (VLC) has wide unlicensed bandwidth, enables communication in radio frequency~(RF) sensitive environments, realizes energy-efficient data transmission, and has the potential to boost the capacity of wireless access networks through spatial reuse. On the other hand, WiFi provides more coverage than VLC and does not suffer from the likelihood of blockage due to the light of sight (LOS) requirement of VLC. In order to take the advantages of both WiFi and VLC, we propose and implement two heterogeneous systems with Internet access. One is the hybrid WiFi-VLC system, utilizing unidirectional VLC channel as downlink and reserving the WiFi back-channel as uplink. The asymmetric solution resolves the optical uplink challenges and benefits from the full-duplex communication based on VLC. To further enhance the robustness and increase throughput, the other system is presented, in which we aggregate WiFi and VLC in parallel by leveraging the bonding technique in Linux operating system. Online experiment results reveal that the hybrid system outperforms the conventional WiFi for the crowded environments in terms of throughput and web page loading time; and also demonstrate the further improved performance of the aggregated system when considering the blocking duration and the distance between access point and user device.
\end{abstract}

\begin{IEEEkeywords}
Hybrid system, heterogeneous network (HetNet), WiFi, visible light communications (VLC), link aggregation.
\end{IEEEkeywords}

\section{Introduction}
The continuous growth in the adoption of mobile devices including
smart phones, tablets, laptops, and now devices on the ``Internet of
Things" is driving an insatiable demand for data access to wireless
networks. This overwhelming demand is rarely met today -- one never
complains about having ``too much bandwidth" or ``too fast service" to
the Internet, especially when considering wireless access. Although
wireless providers are deploying additional access infrastructure by
means of new cells and WiFi end points, the limitation is becoming
overuse of existing RF spectrum. This manifests as contention and
interference and results in an increase in latency and a decrease in
network throughput -- a ``spectrum crunch"~\cite{kavehrad2013optical}. To alleviate this problem, new approaches
to realize larger potential capacity at the wireless link are needed and optical
technologies including visible light communication (VLC) are excellent candidates.

\begin{figure}
\centering
\includegraphics[width=3.0in]{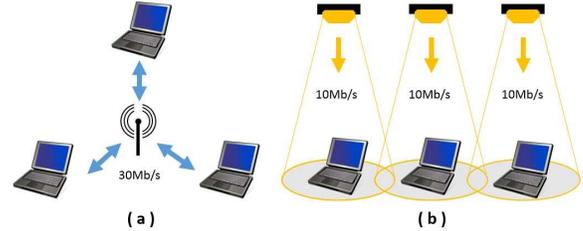}
\caption{Bandwidth density of (a) RF and (b) VLC}
\label{fig_RF_VS_VLC}
\end{figure}

VLC technology provided with LED devices is characterized by high area spectral efficiency,
unlicensed wide bandwidth, high security and dual-use nature
\cite{kahn1997wireless}. For example, Fig.~\ref{fig_RF_VS_VLC} shows
how VLC can reuse spectrum efficiently in a small area. Case a) shows
a WiFi channel in which three users share a 30Mb/s bandwidth, compared
to Case b), a VLC-enabled environment, in which three users utilize
individual 10Mb/s VLC channels. Although the total bandwidth allocated
to the three users are the same in two cases, the outcome aggregated
throughput of Case b) could be better than that of Case a), due to the
contention effect on RF channel as we will see later. As a complementary
approach to the existing wireless RF solutions, VLC is poised to overcome the
crowded radio spectrum in highly-localized systems and become a
promising broadband wireless access candidate to resolve the ``spectrum~crunch".

LED-based indoor VLC has attracted great attention in recent years due
to its innate physical properties including energy efficiency and
lower operational cost compared to conventional incandescent and
fluorescent lighting \cite{komine2004fundamental}. Current research on
VLC focuses mainly on physical (PHY) layer techniques such as dimming
support, flicker mitigation, and advanced modulation schemes
\cite{rajagopal2012ieee}. These efforts seek to achieve the possible highest
data rates. However, higher-level networking challenges must be
addressed to enable interoperability in any practical network
deployment
\cite{rahaim2011hybrid,lee2013performance,chowdhury2013energy,huang2013design}.

Under a dual-use model, VLC is realized by overhead lighting -- lights
serve to provide lighting and also data access. However, providing an
uplink in such a system is challenging due to potential energy
limitations of mobile devices (that do not need to produce
light for illumination) and potential glare from the produced light.
In RF-sensitive and high-security applications, an optical uplink is
possible with relatively high transmission
speed~\cite{schmid2013led}. However, in most RF-insensitive places
such as homes, schools, offices, and supermarkets, an optical uplink
is more difficult to be justified. Mobile devices~(e.g., labtops, smart
phones, tablets) are energy-constrained. Equipping these devices with
a power-hungry light source is impractical. To be efficient, VLC
uplinks will need to use narrow beam widths which lead to challenges
due to device motion and orientation with respect to fixed uplink
receivers. Finally, VLC uplinks can produce glare which is
uncomfortable to and undesirable for human users. Thus VLC remains a
strong contender for the downlink channel but is better if
complemented with an alternative uplink technology.

Alternative heterogeneous schemes, such as VLC and infrared
\cite{langer2007recent}, have been investigated by researchers in
order to resolve the VLC uplink problem at the PHY layer. However, to
make these approaches practical for networking, we still need to address challenges
in realizing upper layer protocols when such an asymmetric model is
adopted. Moreover, the ubiquitous nature of WiFi with its
omnidirectional characteristic can be readily exploited as an
uplink, especially if the use of VLC reduces congestion on the RF downlink
as a heterogeneous network.

\begin{figure}
\centering
\includegraphics[width=2.5in]{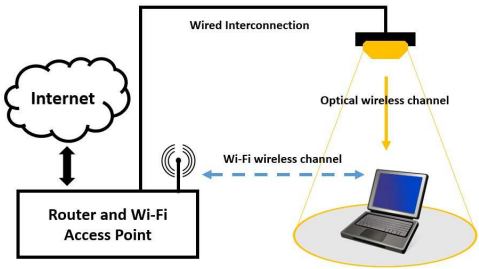}
\caption{Proposed hybrid WiFi and VLC network model}
\label{fig_system_outline}
\end{figure}

In this paper, we propose and implement a practical hybrid system
comprised of typical IEEE 802.11 a/b/g/n technology and a VLC link, in
which the unidirectional VLC channel is exploited to supplement the
conventional downlink RF channel. Such a system was proposed and
theoretically examined in~\cite{rahaim2011hybrid}. Fig.~\ref{fig_system_outline} shows the basic
configuration of this heterogeneous network.  Such a system not only
alleviates congestion caused by WiFi access contention, but also
resolves the potential problems of uplink transmission in VLC
networking. For more information and also the videos of experiments,
please refer to our website\footnote{http://web.njit.edu/$\sim$abdallah/VLC/.}.

To further exploit the potential available resources of WiFi and VLC, we extend
our investigation to the aggregation of multiple wireless interfaces. Although the
hybrid solution alleviates congestion at the WiFi access point, the maximum
achievable download data rate of this system is still limited by the single
VLC link. In environments where VLC hotspots are deployed pervasively, we aim to utilize any available RF resources to supplement VLC links and provide additional capacity to devices requiring higher throughput. Referring to~\cite{ramaboli2012bandwidth},
we can benefit from the bandwidth aggregation for not only the improved throughput, but also
the reliable packet delivery, load balancing and low cost capacity increase.
It is also shown that many efforts have been spent on the bandwidth aggregation
at different layers of the network protocol stack. Since our main objective is to implement
the aggregation with modification to the clients only without affecting the server part,
we focus our attention on the data link layer aggregation.

\begin{figure}
\centering
\includegraphics[width=2.5in]{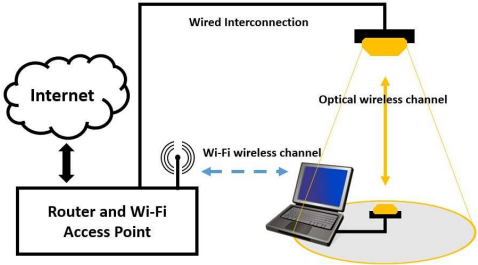}
\caption{Proposed aggregated WiFi and VLC network model}
\label{fig_bonding_system_outline}
\end{figure}

In this paper, we utilize link aggregation through the use of two full-duplex
wireless connections. Both the bi-directional WiFi and VLC links are fully utilized to
improve the achievable throughput and provide a more robust network connectivity. Fig.~\ref{fig_bonding_system_outline}
depicts the aggregated network. With combining multiple wireless access technologies,
the system takes the advantages of both communication techniques.

The main contributions of the work are the following:
\begin{itemize}
\item The design and implementation of an asymmetric system comprised of WiFi uplink and VLC downlink to increase overall network capacity with multiple users.
\item The design and implementation of an aggregated system that simultaneously activates both WiFi and VLC connections providing high throughput and more reliable data transmission.
\item Analysis and real experimentation on our testbed to evaluate the network performance of two different systems under interactive web browser traffic and TCP throughput with different levels of congestion.
\end{itemize}

The paper is organized as follows. Section \ref{sec2} reviews related
work on hybrid WiFi and VLC systems and data link layer aggregation. Section \ref{sec3}
describes the designed asymmetric system in detail including the router
reconfiguration, packet capture and retransmission; and most
significantly, the network-level operating system adaptations to
realize the asymmetric protocols. We also demonstrate the detailed implementation of the aggregated system
based on bonding driver in Linux operating system. Section \ref{sec4} provides analysis and
experimental results demonstrating the benefit of the proposed hybrid
and aggregated systems. Section \ref{sec5} concludes the paper.

\section{Related Work}\label{sec2}
Early work on hybrid systems integrating RF and VLC are based on
simulation and analysis \cite{rahaim2011hybrid,lee2013performance,chowdhury2013energy,huang2013design}.
To the best of our knowledge, none of these hybrid systems were able to develop
practical system implementation yielding functional IP-based communication
supporting web browsing or other Internet access functionalities.


A model for integrating WiFi and VLC has previously been
proposed but not implemented \cite{rahaim2011hybrid}. In this model, downlink VLC
channels are proposed to supplement an existing RF channel. Handover
techniques are defined for resolving discontinuities due to mobility
and specifically to transfer between a symmetric RF link and the
asymmetric VLC-RF link as a device transits an indoor space. In this
prior work, the primary contributions are simulation and analysis of
the downlink channel under the assumption of a reliable RF uplink.

Device cost and energy consumption relative to data throughput have
been investigated for a hybrid VLC system~\cite{lee2013performance}. The authors show advantages for an RF
uplink compared to a VLC uplink, but primarily as related to energy
cost for transmission. This work motivates us to study the hybrid system
that replaces the energy-expensive uplink with~RF.

Energy-efficient connectivity for a hybrid radio-optical wireless
systems has also been investigated in \cite{chowdhury2013energy}. In this
work the authors show via simulation that connectivity and energy consumption depend
on user device density, coverage range ratio between single-hop and
multi-hop, relay probabilities, and mobility of the user. Although the
proposed WLAN-VLC network model shows the positive impact of a
hybrid system, the approach they used relies on an ideal scenario that entails prior
assumptions.

Room division multiplexing (RDM) has been demonstrated under a hybrid
VLC network model \cite{huang2013design}. The core component of this
hybrid system is the VLC network coordinator, which is responsible for
RDM-based service division and distribution as well as for providing
bidirectional interfaces between the outdoor and indoor communication
infrastructure, especially the indoor interfaces for uplink WiFi
access and downlink LED lamps. This work, however, does not appear to
extend to full implementation of the network protocol stack nor
implementation in the system kernel.  Finally, the work is evaluated
by waveform measurement without the signal processing and demodulation
required for practical use.

Each of the aforementioned works focuses on simulation analysis
without implementation of the full end-to-end system required to
provide evaluation at the application layer. In contrast, we implement a practical hybrid WiFi-VLC wireless
system, which enables the typical Internet access connection between client and
server {\em without any reconfiguration at the server side}. Data packets
generated by user applications are transmitted through WiFi and
requested data from the server is received via the VLC interface.

Early efforts on the data link layer aggregation were also based on the simulations \cite{koudouridis2005generic,koudouridis2005switched,yaver2009performance,kim2008mac,kim2010feedback,kim2012splitting}. To the best of our knowledge, none of the
above work evaluates the performance of their presented policy based on real experiments.

The concept of generic link layer (GLL) that aggregates multiple radio access at the radio level is introduced in \cite{koudouridis2005generic}. The GLL enables multi-radio transmission diversity (MRTD), which transmits a traffic flow sequentially or parallel over aggregated radio access technologies. The significant functions in GLL include access selection schedule, performance monitoring and flow error control. GLL blindly assigns the traffic to different interfaces. Based on the GLL concept, the authors further study the switched MRTD in \cite{koudouridis2005switched}. Based on the measured throughput of each radio access technology, the work sorts the interfaces in descending order of their available throughput. The traffic flows are distributed to the interfaces in the descending order.  In \cite{yaver2009performance}, another factor, round trip time (RTT) is utilized to schedule the traffic through the interfaces. In \cite{koudouridis2005switched} and \cite{yaver2009performance}, the switched MRTD is presented as an adaptive mode of the original MRTD. However, with similar available throughput over the interfaces, the proposed scheme produces inevitable redundant switching overheads.

A cognitive convergence layer (CCL), which is similar to GLL is introduced in \cite{kim2008mac}. The authors propose a logic link layer interface for managing the traffic flows through real link layer interfaces. A traffic distribution policy is implemented at the sender and a reorder buffer is added at the receiver. A function of link transmission time (LTT) is used to estimate the links capacities. The CCL-based data link aggregation is further investigated in \cite{kim2010feedback}. The authors use the link delay as a criterion to determine the data traffic distribution over tightly-coupled WiMAX and WiFi network. In \cite{kim2012splitting}, the measurement of channel occupancy time for a single packet transmission is used as the decision metric. The air-time cost based scheme is reported as a more adaptive link aggregation schedule than the RTT-based policy.

All of the above-mentioned data link layer aggregation works focus on the network parameters, which are used for determining the link selection. None of them provided implementation of their proposed schemes or experimental analysis. In our work, we set up the testbed based on the bonding driver of Linux OS and evaluate the TCP throughput as well as the user experience for web browsing.

\section{System Model}\label{sec3}
\begin{figure}
\centering
\includegraphics[width=8cm]{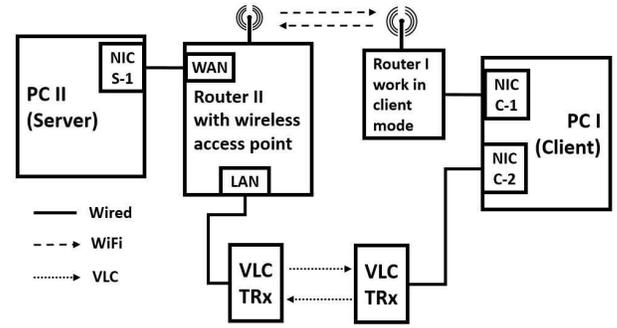}
\caption{Aggregated system architecture}
\label{fig_Aggregated_system_design}
\end{figure}

In this section, we illustrate the models of two heterogeneous systems: i) Hybrid WiFi-VLC system (Fig.~\ref{fig_system_design}); ii)~Aggregated WiFi-VLC system (Fig.~\ref{fig_Aggregated_system_design}).

\subsection{Hybrid System}

\textbf{Challenges}: The primary challenges of designing an asymmetric system are as follows:

1) Typically, uplink and downlink data streams based on WiFi connection between the client and the server flow through the same routing path. In order to redirect the data flow downloaded from the server to the client to the VLC hotspot, an intermediate coordinator is needed to break the conventional downlink data delivery and forward the data packets to the VLC hotspot. However, this process may generate inevitable redundancy due to the need of extra devices to perform data redirection. In Fig. \ref{fig_system_design}, we aim to assign the infrastructural router as the redirecting node, forwarding downlink traffic to PC I and simultaneously providing PC II with an uplink wireless access point.

2) A typical small office home office (SOHO) wireless router has one wide area network (WAN) port and multiple local network (LAN) Ethernet ports. Terminals connected to the router through either wired or wireless links belong to the same sub-network. This router serves as an edge router with a gateway IP address. Intuitively, we might be able to activate the routing function on PC I in Fig. \ref{fig_system_design}, in order to directly forward the data packets from the router to PC II. However, due to the OS kernel built-redirecting function, the simple forwarding method based on the routing function may not be useful. Since the destination IP address of the data packets that arrive at the network interface card (NIC) A-1 is actually the IP address of NIC B-1, the packets will be redirected back to PC II through the router instead of the VLC link, if the forwarding function of PC I is activated.

3) In a typical TCP connection, the client initiates a three-phase handshake process with the server. According to the OSI model, the client first generates a SYN data segment at the application layer. After that, the data segment is encapsulated with IP headers at the network layer before being sent out through the NIC. Since the client starts listening to the socket with the same TCP port and IP address as that used during encapsulation, the following problem occurs. If the packets from the server are received from a different NIC with different socket information, they may not be selected by the application that initiated the TCP connection. In Fig.~\ref{fig_system_design}, the requests are transmitted through NIC B-1 while the responses are received from NIC B-2. The above problem is encountered in the asymmetric system in Fig.~\ref{fig_system_design}.

\begin{table}
\centering
\caption{An example of static routing table}
\begin{tabular}{|c|c|c|c|} \hline
Dst IP&Subnet Mask&Next hop&Metric\\ \hline
192.168.1.100&255.255.255.255&192.168.1.200&2\\ \hline
\end{tabular}
\end{table}

\textbf{System Design}: Fig. \ref{fig_system_design} demonstrates the hybrid system model for indoor Internet access. The system consists of a downlink VLC channel and an uplink WiFi channel. To resolve the challenges mentioned earlier, three procedures (each one of these procedures corresponds to one of the above challenges) need to be performed as follows:

\begin{figure}
\centering
\includegraphics[width=8cm]{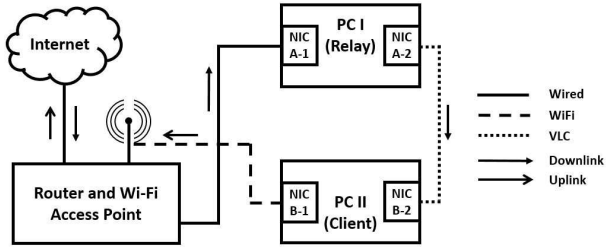}
\caption{Hybrid system architecture}
\label{fig_system_design}
\end{figure}

\begin{algorithm}
\caption{Pseudo code of socket program}\label{algorithm}
\begin{algorithmic}[1]
\REQUIRE~~\\
Define BufferSize MTU;\\
Set socket s for frames capture;\\
Set socket d for frames retransmission;\\
Bind socket s to NIC A-1;\\
Bind socket d to NIC A-2;\\
\ENSURE~~\\
\WHILE {1}
\STATE Receive frames from socket s and store into buffer msg[BufferSize];
\IF {frame length $>$ MTU}
\STATE Continue;
\ENDIF
\IF {frame desitination IP addr = IP B-1}
\STATE Change dest MAC addr to MAC B-2;
\STATE Change src MAC addr to MAC A-2;
\STATE Change dest IP addr to IP B-2;
\STATE Compute IP checksum;
\STATE Compute TCP checksum;
\STATE Compute UDP checksum;
\STATE Send modified frames to socket d;
\ENDIF
\ENDWHILE
\end{algorithmic}
\end{algorithm}

1) To address the problem mentioned in the first challenge, a static routing table is enabled at the router. Rather than dynamically forwarding IP packets as normal, the router follows a manually-configured routing entry with three items: i)~destination IP address, ii) subnet mask, and iii) next-hop router IP address. Table 1 shows an example of routing IP traffic destined for the 192.168.1.100/24 via the next-hop router with the IPv4 address of 192.168.1.200/24. In the proposed hybrid system, one active static routing rule redirects IP packets destined for NIC B-1 to NIC A-1 instead.

2) After successfully arriving at the relay node (PC I), server IP packets need to be further forwarded to the client through NIC A-2. In the Linux OS, the IP packet forwarding function is enabled by changing the value of {\em ip\_forward} under the path ``/proc/sys/net/ipv4/ip\_forward" from 0 to 1. As mentioned in the challenge (2) earlier, if we activate the forwarding function on PC I, the arrived IP packets will be redirected back to PC~II through NIC A-1 instead of NIC A-2. Therefore, we must set the value of {\em ip\_forward} to 0. Rather than relying on the forwarding function, we utilize the socket programming based on $SOCK\_PACKET$ type \cite{senie2002using}.

The $SOCK\_PACKET$ mechanism in Linux is used to take complete control of the Ethernet interfaces. Due to the capability of capturing frames from the data link layer and placing a pointer which points to the first byte of each frame (the first byte of MAC header), $SOCK\_PACKET$ is suitable for MAC frames capturing and retransmission. Algorithm 1 represents the relaying functionality. In the algorithm, we first define the buffer size according to the maximum transmission unit (MTU) which is a default value in the router. Then two sockets of the $SOCK\_PACKET$ type are created and bound to NIC A-1 and NIC A-2. After the initialization phase of key parameters, the iteration phase, which includes receiving, processing and retransmission, is started. To avoid the alert of no buffer space available for the sending function, the received frame length needs to be checked. If the length is larger than the defined MTU, the frame must be discarded. Also, the destination IP address of the captured frames for efficient relaying is checked. If it is the same as the IP address of NIC B-1, we manually modify packet's MAC and destination IP address. To realize the Internet access, the checksums of IP, TCP and UDP need to be recomputed before sending the packets to the client. This is because the checksum computation includes the destination IP address.

\begin{figure}
\centering
\includegraphics[width=3.5in]{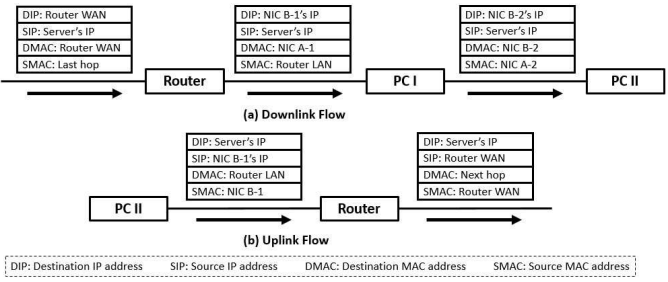}
\caption{Flow of MAC and IP headers of packets between server and client, a) downlink flow and b) uplink flow}
\label{fig_IP_MAC_headers}
\end{figure}

3) As it is mentioned in challenge (3) earlier, the application that initiates the TCP connection to the server will listen to the socket with the IP address of NIC B-1 but not NIC B-2. The OS kernel will not do any action on the packets although they are not filtered by the NIC. A possible solution is changing the destination IP of the packets to the IP address of NIC B-1. However, due to the fact that the destination IP address of the packets is not that of the port they are received from, the packets should be forwarded based on the routing table. To overcome this difficulty, an approach called ``operating system spoofing" is proposed.

{\it Operating System Spoofing:} The basic idea of this approach is to manually make the OS listen to NIC B-2 while transmitting out the packets through NIC B-1. Assume that the IP addresses of NIC B-2 is 192.168.2.100/24 and the default gateway of PC II is 192.168.1.1/24. The default gateway is deleted and a new one within the subnet 192.168.2.0/24 (e.g. 192.168.2.1/24) is added. In addition, an entry in the ARP table on PC II (e.g. arp -s 192.168.2.1 ab:ab:ab:ab:ab:ab) is added. When these two steps are completed, PC II will believe that there is a next-hop gateway connected to NIC B-2 even though this gateway does not exist. With the ARP spoofing, all packets generated at the application layer on PC II are forwarded to NIC B-2 and stop there because of the physical layer blocking. The most significant point at this moment is that the applications are listening to the socket with the IP address of NIC B-2.

After the configuration of the routing and ARP tables, a socket program, that implements packets copying, headers modification and retransmission, is run. Similar to the program run on PC I, the socket of type $SOCK\_PACKET$ is used to capture the packets flowing through the device driver layer of NIC B-2. With the returned pointer, the source IP and MAC addresses of the copied packets to the IP and MAC addresses of NIC A-1 are altered. Also, we change the destination MAC address to router's LAN MAC address. The checksums of IP, TCP and UDP need to be recomputed. After all the modifications of the IP and MAC headers are completed, the packets are sent to the router through NIC B-1. From the router's point of view, the client's IP is the IP address of NIC B-1. However, from the client's point of view, it connects to the Internet with the IP address of NIC B-2. Fig.~\ref{fig_IP_MAC_headers} reveals the variation of IP and MAC headers.

Note that, compared to the previously published work \cite{shaoindoor}, the achievable network throughput of the proposed hybrid system has been enhanced. For the packets captured on PC II, a selective condition before sending them to the router is added. Only the packets with a source IP same as the IP address of NIC B-2 are processed by the program. Since the capture function returns the packets not only in the transmitting buffer of NIC B-2 but also in the receiving buffer, the added condition saves half of the redundant processing time in the unmodified program.

\subsection{Aggregated System}
Fig.~\ref{fig_Aggregated_system_design} demonstrates the link aggregation system model including the full-duplex WiFi and VLC connections. Since the bonding driver provided by Linux operating system is only capable of aggregating Ethernet interfaces, both NIC C-1 and NIC C-2 of PC I (client) are Ethernet cards. The wireless Router I working in client mode, is connected to PC I through Ethernet cable and connected to Router II through bi-directional WiFi link. The other network connection between the PC I and the Router II is a bi-directional VLC link, which is established by two VLC transceivers. The bi-directional VLC link will be described in section~\ref{sec5}. The two NICs on PC I are in the same subnet and the IP address of Router II LAN is used as the IP address of the gateway of PC I, when the aggregation configuration is not enabled. On the right side, PC II (server) is connected to the WAN port of Router II. Thus Router II acts as an intermediate node between two different networks.

To implement the data link layer aggregation, we utilize the Linux Ethernet bond driver \cite{davis2011linux}. The driver provides a method to aggregate multiple Ethernet interfaces into a single logical ``bond" interface. The behavior of the artificial interface depends on the selected mode. There are seven modes in bonding configuration: 0) Balance-rr; 1) Active-backup; 2)~Balance-xor; 3) Broadcast; 4) 802.3ad; 5) Balance-tlb; 6) Balance-alb. Modes 0, 2 and 4 require extra switch support, which conflict with our objective of not having extra devices. Under Mode 1, only one slave in the bond is active. The other slave becomes active if and only if the active slave fails. Thus, the maximum throughput that can be achieved in this mode can not exceed the highest one of the slaves. In Mode 3, the same packets are sent through all slave interfaces, in order to provide fault tolerance. Therefore, only Mode 5 and Mode~6 are left to choose from. For the goal of achieving higher aggregated throughput without any modification to the server side or even to the next hop of the client, Mode 6 is the best option.

Mode 6 (adaptive load balancing) contains Mode 5 (adaptive transmit load balancing). Also, Mode 6 integrates the receive load balancing for the IPv4 traffic and does not require any switch support. Load balancing at the receiver is achieved by ARP negotiation. The bond driver intercepts the ARP response sent from the host and changes the source MAC address to the unique MAC address of one of the slaves. It enables the peers to use a different MAC address for communication. Typically, all the slave ports will receive the broadcast ARP requests from the router. The bond driver module intercepts all the ARP responses sent from the client, and computes the corresponding port that the client expects to receive data from. Then the driver modifies the source MAC address of the ARP response to the MAC address of the corresponding port. The destination MAC address is kept the same as the MAC address of the router LAN. Note that each port can send the ARP response not only with its own MAC address but also with the MAC address of the other slave port. The received data traffic can be load balanced in either way. When the client sends the ARP request, bonding driver copies and saves the IP information of the router. When the ARP reply from the router arrives, the bond driver extracts the MAC address of the router and sends an ARP response to one of the slave ports (this process is the same as the received load balancing process mentioned above). One potential problem in ARP negotiation is that when the client sends out the ARP request, it uses the MAC address of the logic bonding interface. Thus, after the router learns this MAC address, the downlink traffic from the server flows through the corresponding slave port which may not be the intended one. This problem can be addressed by sending the updated ARP response. The client sends the ARP responses to all slave ports and each response contains the unique MAC address of each slave port. Thus, the downlink traffic from the server are redistributed. The receive load \cite{davis2011linux} is allocated orderly from the slave with widest bandwidth.

In Fig.~\ref{fig_Aggregated_system_design}, when Router II broadcasts the ARP request to PC~I, PC~I typically sends back an ARP response with the bonding MAC address and the bonding IP address. However, under the adaptive load balancing mode, the bonding driver intercepts the APR response and changes the source MAC address to that of one of the slaves (e.g. NIC C-1). Thus, when the router receives the ARP response, it refreshes its ARP cache with a new entry (Bond IP: NIC C-1 MAC). To increase the total bandwidth of the client, PC I sends back the ARP response with the NIC C-2 MAC address when the capacity of NIC C-1 is exhausted. After receiving the new ARP response, the router updates its ARP cache (Bond IP: NIC C-2 MAC).

\subsection{Analysis}
``Spectrum crunch" \cite{kavehrad2013optical} is a challenging problem in WiFi networks. Due to the limited bandwidth, although the efficiency of the spectrum utilization has been highly improved, the degradation of network throughput caused by the growing number of WiFi users and other devices operating in the 2.4 and 5 GHz bands is still inevitable. Also, the network delay could be much larger when the number of users in the same WiFi access point increases. Because of the CSMA/CA mechanism defined in 802.11 standards \cite{bianchi1996performance}, the average back-off time is unavoidably increased when there exists more mobile users located in the same WiFi coverage. To circumvent these unexpected user experiences, the hybrid WiFi-VLC system is presented. Although the uplink WiFi may be in contention with other WiFi users, the downlink VLC is an independent data communication channel from the WiFi channels. In our daily life, downloading happens much more frequently than uploading; hence having an undisturbed VLC download channel may provide a satisfactory Internet surfing experience. Regarding the increasing demand for downlink bandwidth, the hybrid VLC user does not need to compete with other WiFi users for the RF spectrum. Therefore, the network delay will be reduced.

In some specific scenarios where the uplink VLC is allowable, aggregating both bi-directional WiFi and VLC links can provide more available bandwidth. Taking the shortages of VLC into account, the aggregated system is more robust than the hybrid one. As the distance between the front-ends of WiFi and VLC is increasing or the channel blocking duration is increasing, the performance of VLC channel may drop quickly. However, these two factors are not influential in WiFi communication when considering the short distance for indoor links. Aggregation may not only achieve higher average throughput, but also maintain the advantages of both WiFi and VLC.

\section{Experiments}\label{sec4}

\subsection{VLC front-ends and performance of the single VLC link}
\subsubsection{VLC front-ends}
\begin{figure}
\centering
\includegraphics[width=7.0cm]{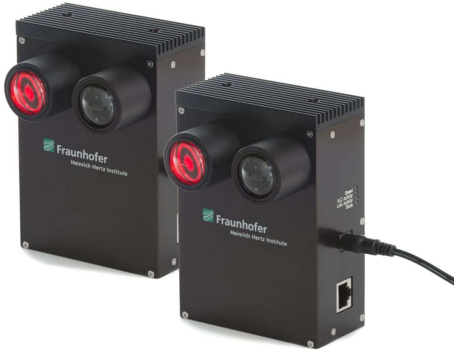}
\caption{VLC front-ends}
\label{fig_VLC_frontends}
\end{figure}

Similar to RF communication, the capabilities of VLC strongly depend on the analog front-ends such as power amplifiers and antennas. Also, the optical source and photodetector have a great effect on the performance of VLC. Our proposed systems, coexisting WiFi and VLC, require high-speed front-ends with PHY and MAC layer implementation. The PC-LED contains a blue light source and enveloped by a yellow phosphor to produce white light. This low-cost LED provides only narrow modulation bandwidth caused by slow response time of the phosphorescent material. Our collaborators in Fraunhofer-HHI have recently developed a small form factor current driver using an off-the shelf high power white LED. The VLC receiver consists of a transimpedance amplifier (TIA) and a commercially available high-speed Si-PIN photodiode (PD). It is reported that the modulation bandwidth has been improved from 3-7 MHz to 20 MHz. This enhancement is realized by reducing the effect of phosphorescent portion in the optical spectrum with the aid of a blue filter at the receiver end. The new VLC transceivers shown in Fig.~\ref{fig_VLC_frontends} has significantly increased the modulation bandwidth to above 100 MHz by means of precise impedance matching between the LED and the high-power analog driver as well as between the PD and the low-noise amplifier followed by. Fraunhofer-HHI has currently implemented an available bi-directional VLC link with a 150 MHz analog transmitter, a 100 MHz analog receiver and a 70 MHz orthogonal frequency-division multiplexing~(OFDM) baseband processor. As shown in Fig~\ref{fig_VLC_frontends}, each device comprises an external power supply and a 1 Gbps Ethernet port using RJ45 standard. Two modules (transceivers) are operating like an Ethernet bridge allowing the transfer of all kinds of data. Altogether, a gross and net data rate of 500 and 260 Mbps are possible with one-way latency of around 10 ms.

\begin{figure}
\centering
\includegraphics[width=8.9cm]{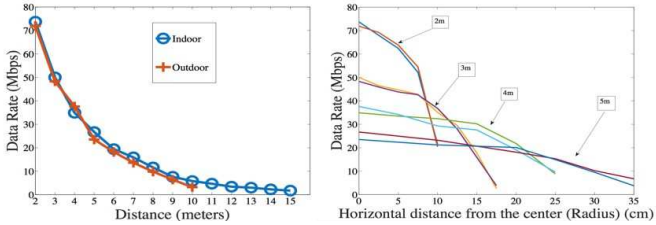}
\caption{Throughput vs. vertical and horizontal distance between VLC transceivers}
\label{fig_vertical_horizontal}
\end{figure}

\subsubsection{Performance of indoor and outdoor VLC links}
Based on the VLC front-ends developed by our collaborators at Fraunhofer-HHI, we conduct experiments indoor and outdoor to evaluate the average throughput of the VLC link. For indoor experiment, we measure the throughput within the range of 2-15 meters while the range is 2-10 meters for outdoor experiment. Note that the distance here represents the vertical distance between two transceivers and the horizontal distance is equal to 0. We also measure the throughput at various horizontal distance within the coverage of the VLC source. All the measurement results are averaged over 100 runs. Each run is a throughput test and spends 5 seconds. Fig.~\ref{fig_vertical_horizontal} (left) represents the average throughput vs. the vertical distance. We achieve around 74 Mbps at 2 meters and the throughput drops to about 25 Mbps at 5 meters. Note that the vertical distance for most indoor applications is in this range. Also note that the throughput results here with short vertical distance are higher than the typical throughput that can be achieved by using WiFi with 2.4 GHz channel conformed to standard 802.11 b/g/n. As we can observe, the throughput results for the outdoor experiment are very similar to those for indoor experiment. Note that the outdoor environment can represent the extreme condition that happens indoor, with windows open and extensive lighting. The results prove that the VLC front-ends are suitable for the extreme indoor case. Fig.~\ref{fig_vertical_horizontal} (right) shows the average throughput when we vary the horizontal distance within the vertical distance from 2m to 5m.

Note that the data rates measured here are associated with both, a customized configuration of the VLC front-ends and a different metric used, compared to \cite{langeroptoelectronics} and \cite{grobe2013high}. In particular, different LEDs are used (downlink is white LED while uplink is infrared, instead of red and blue, respectively). Red LED power is higher in the downlink and the PD is more sensitive to reddish colors. Due to eye safety, infrared power is more limited in the uplink than for blue light. Finally, the results in the present paper represent the net data rate measured on the application layer in the OSI model while our previously published data refer to the gross data rate at the physical layer, as it is measured at the baseband processing chipset.

\subsection{Testbed of the hybrid system}
In the testbed of our proposed hybrid system, we have two PCs with Linux OSs (Ubuntu 12.0.4 LTS), a NETGEAR Wireless Dual band Gigabit Router WNDR4500, and two VLC transceivers provided by Fraunhofer-HHI. The PC which performs relaying functions is equipped with two Ethernet cards: Inter Corporation 82579LM and 82574L Gigabit Ethernet Controllers. The client PC is equipped with one wireless card and one Ethernet card: a Boradcom 802.11n Network Adapter and a Broadcom NetXtreme Gigabit Ethernet controller respectively. All the NICs support 10/100/1000M speed.

Regarding the network configuration, router's LAN IP address is set to 192.168.1.1/24 as default. Referring to Fig.~\ref{fig_system_design}, the IP addresses of NIC A-1, NIC B-1, NIC A-2 and NIC B-2 are manually configured as 192.168.1.200/24, 192.168.1.100/24, 192.168.2.200/24 and 192.168.2.100/24 respectively. The IPv4 routing table in client PC is shown in Table 2. And an additional entry in client's ARP table is added by typing ``arp -s 192.168.2.1 ab:ab:ab:ab:ab:ab" in command window with root privilege.

For the VLC unidirectional link setup, since the VLC devices provided by Fraunhofer-HHI are both transceivers, we manually turn off the forwarding function in PC I (relay), in order to construct a network-level unidirectional VLC channel. Regarding the VLC connection establishment, there are Ethernet ports on the VLC transceivers, therefore, simply connecting the transceivers to the PCs with Ethernet cable constructs the wireless VLC link.

\subsection{Testbed of the aggregated system}
\begin{figure}
\centering
\includegraphics[width=5cm]{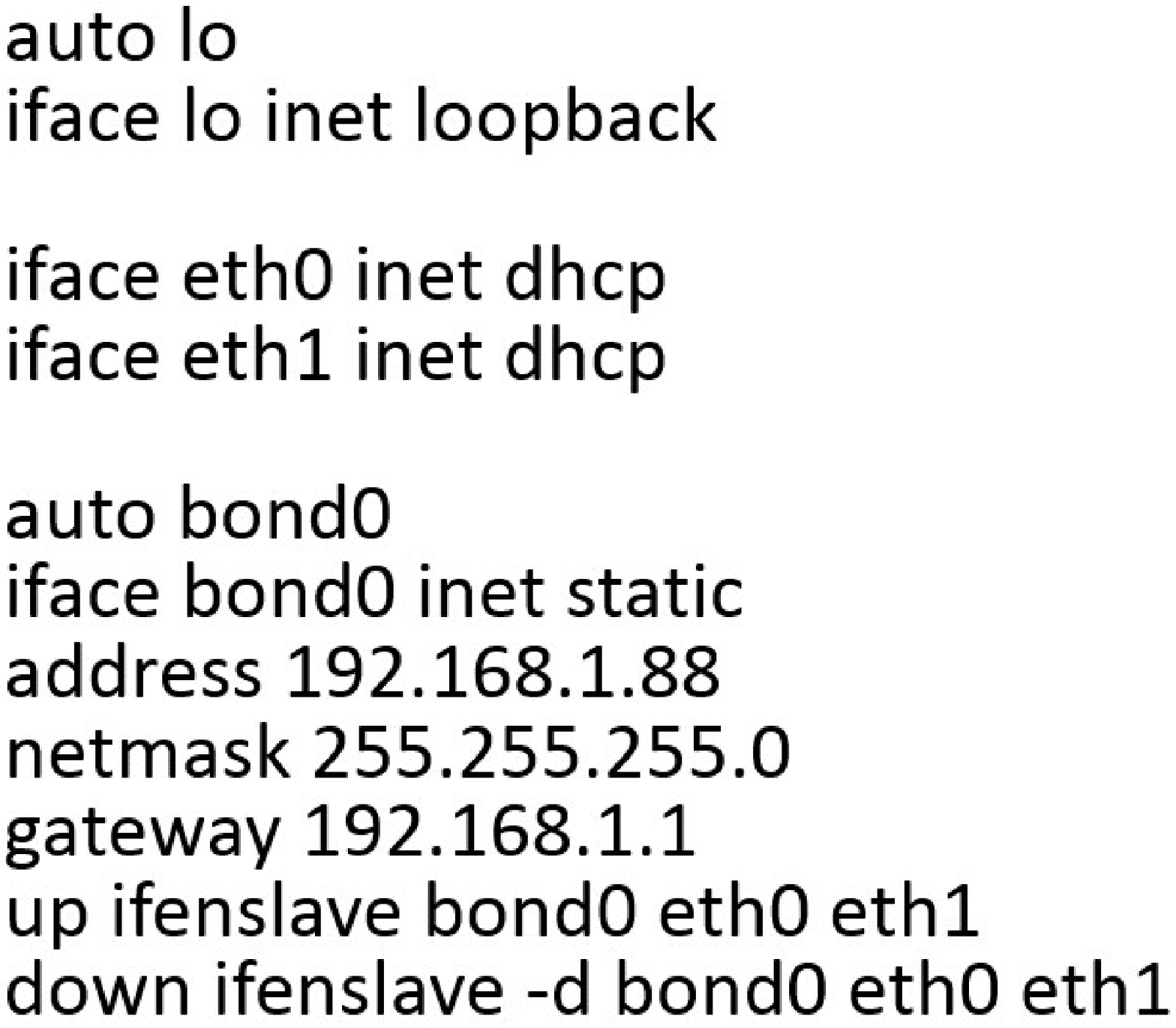}
\caption{Bonding configuration}
\label{fig_bonding_commands}
\end{figure}

To implement the aggregation of two wireless links, we use one PC with Linux OSs (Ubuntu 12.0.4 LTS), a NETGEAR Wireless Dual band Gigabit Router WNDR4500, a client mode  TP-LINK wireless router 150 Mbps TL-WR702N, and also the two VLC transceivers. The client PC is equipped with two Ethernet cards: Inter Corporation 82579LM and 82574L Gigabit Ethernet Controllers. Both the NICs support 10/100/1000M speed.
Regarding the network configuration, in order to construct the aggregated system, we add one line to /etc/modules: ``bonding mode=6". After that, we modify the /etc/network/interfaces. We change the eth0 and eth1 to auto DHCP and also add the static logic bond0. The detailed commands are shown in Fig.~\ref{fig_bonding_commands}.

\begin{figure}
\centering
\includegraphics[width=7.8cm]{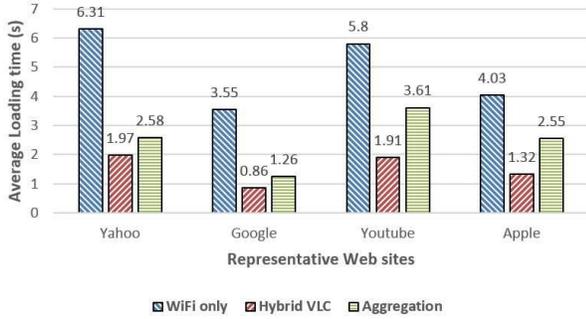}
\caption{Loading time in web browsing}
\label{fig_loadtime}
\end{figure}

\begin{figure}
\centering
\includegraphics[width=7.8cm]{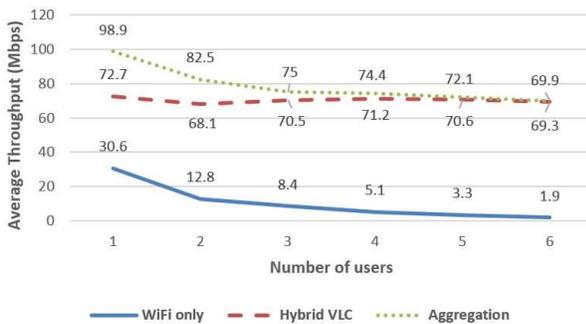}
\caption{Throughput vs. Number of contenders}
\label{fig_throughput}
\end{figure}

\begin{figure}
\centering
\includegraphics[width=7.8cm]{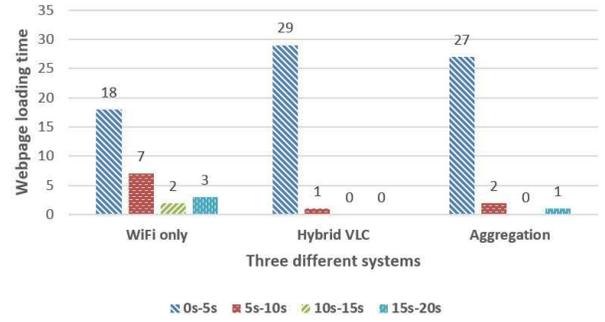}
\caption{Statistic of yahoo homepage loading time}
\label{fig_statistic}
\end{figure}

\begin{figure}
\centering
\includegraphics[width=7.8cm]{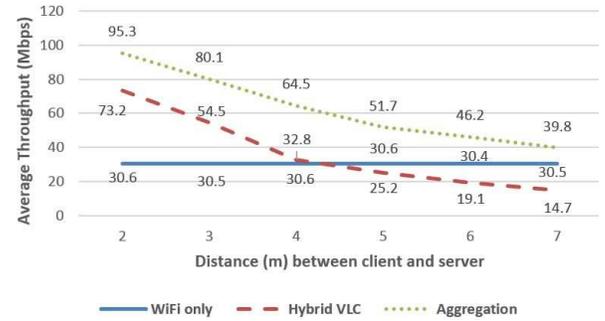}
\caption{Throughput vs. Distance between tx and rx}
\label{fig_distance}
\end{figure}

\begin{figure}
\centering
\includegraphics[width=7.8cm]{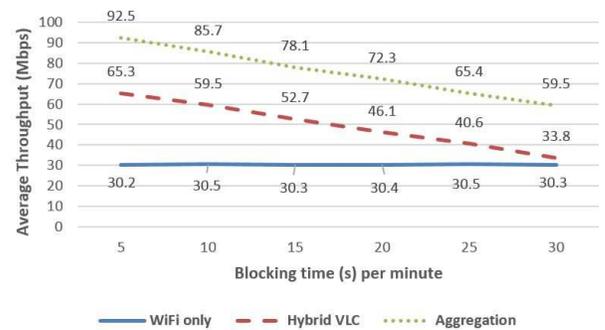}
\caption{Throughput vs. Block duration}
\label{fig_block}
\end{figure}

\begin{table}
\centering
\caption{Routing table of client}
\begin{tabular}{|c|c|c|c|c|c|} \hline
Destination&Gateway&Genmask&Flags&Metric&Interface\\ \hline
0.0.0.0&192.168.2.1&0.0.0.0&UG&0&eth0\\ \hline
169.254.0.0&0.0.0.0&255.255.0.0&U&1000&eth1\\ \hline
192.168.2.0&0.0.0.0&255.255.255.0&U&2&eth0\\ \hline
\end{tabular}
\end{table}

\subsection{Results and Analysis}
Iperf \cite{tirumala2005iperf} is a pervasively used tool to measure the network performance. Due the limited bandwidth allocated by the ISP, we setup an internal network with a client, a server and a router, for the purpose of measuring the achievable bandwidth of system. We install iperf on both the client and the server to test the average TCP throughput in all the experiments except the test of web page loading time. All throughput results are averaged over 100 runs. Each run spends 5 seconds.

In Fig.~\ref{fig_throughput}, we show the average of TCP throughput achieved by the three different systems(the two systems described in section \ref{sec3} in addition to a system that uses WiFi only). The distance between the transmitter and the receiver is 2m for both WiFi and VLC. As the number of contending WiFi users is increasing, we test the average TCP throughput obtained by a selected user who uses the three different systems. We set the wireless mode of the router to ``up to 54 Mbps", hence the TCP throughput achieved by the single WiFi user is around 30 Mbps. As the number of users is increasing, the performance of WiFi declines sharply. In contrast, since the contending WiFi users only contend with the uplink wireless channel of hybrid VLC, the throughput of the hybrid VLC user is around 70 Mbps, regardless of the number of users, which is 5 times higher than the WiFi only users when the number of users increases to 6. Also, we can observe that the higher throughput achieved by the aggregated system, which is represented by the dot line, is almost the aggregated value of the throughput of WiFi only and hybrid VLC. To satisfy the specific users who need extremely high downloading data rate, the aggregated system can be utilized. Note that compared to our previously published work \cite{shaoindoor}, the average throughput is enhanced from 150 Kbps (limited by the software-defined-VLC) to 70 Mbps, which is increased by around 477 times.

In addition to the throughput measurement, we also evaluate the loading time of web browsing by selecting several representative web sites. Pingdom\footnote{http://tools.pingdom.com/fpt/.} online website speed test is used to estimate the loading time of the webpages. As shown in Fig.~\ref{fig_loadtime}, we investigate the completion time of the home webpages of yahoo, google, youtube and apple on one client located in a network comprised of 10 clients (the other 9 WiFi users are downloading a 1GB file through the same access point). We can observe that the shortest loading time occurs when the selected user chooses the hybrid VLC system. Theoretically speaking, only the WiFi channel could be congested by the other WiFi users, it is reasonable that the loading time of aggregated system is shorter than the WiFi only but longer than the hybrid VLC. Therefore, based on our implementation, the network performance of the aggregated system is not always better than the hybrid system. This is because the condition in section \ref{sec4} are not always satisfied. Our final results of loading time are averaged over 30 runs. Each run is a web page loading time test via Pingdom website. The statistics of the yahoo's homepage loading time are shown in Fig.~\ref{fig_statistic}. Most of the loading time in hybrid and aggregated systems are distributed within the range of 0s-5s. However, 3 tests results are distributed in the range of 15s-20s in WiFi only system. With the WiFi access point congested, the level of network delay may be highly degraded due to the increased back-off penalty.

The short range factor of VLC is inevitable, hence we vary the distance between the transmitter and the receiver for both WiFi and VLC, to measure the average TCP throughput. The experiments results are shown in Fig.~\ref{fig_distance}. As the distance is increasing, the achievable throughput of hybrid VLC user is decreasing quickly. However, the WiFi only system has a stable network performance. The throughput of the WiFi only exceeds the hybrid VLC when the distance is increased to 4.1m. To some extent, the aggregated system is capable of providing a throughput with the lowest bound that is higher than the WiFi only. Therefore, the aggregation technique improves the integrated data rate and offers users with more reliable communication.

Due to the particularly small wavelength of the visible light, VLC channel can be easily blocked by tiny objects. Regarding the irregular movement of mobile devices, the channel blockage duration becomes a considerable factor when we evaluate the network performance of VLC. We vary the blocking time from 5s to 30s per minute and test the average TCP throughput achieved by the three systems. Fig.~\ref{fig_block} demonstrates the experiments results. Even if the blocking time in VLC channel is increased up to 30s per minute, the achievable TCP throughput is still higher than the WiFi only system. Therefore, compared to the distance, the blocking duration may be less influential. Additionally, although the throughput of aggregated system is also decreased, it would not be lower than that of the WiFi only system. The blockage duration experiment further proves the robustness of the aggregated system.

\section{Conclusion and future work}\label{sec5}
In this paper, we evaluate two heterogeneous systems incorporating WiFi and VLC. Our goal is to provide a proof of concept for the coexistence between these two communication bands. Within a short distance between the transmitter and the receiver, the hybrid VLC could perform much better than the WiFi system in the crowded wireless environment. As a complementary technique, VLC deserves further investigation. However, on one hand, WiFi infrastructures are prevalent and highly acceptable by most consumers; on the other hand, WiFi may outperform VLC in case of long distance data transmission or the existence of obstacles. Therefore, we conclude that the aggregation between WiFi and VLC is worthy of further study, to effectively utilize the aggregated bandwidth and to lower the network delay.

For future work, we intend to apply aggregation on the hybrid VLC system. To resolve the challenges of optical uplink in our implemented aggregated system, an approach that integrates the symmetric WiFi only link and the asymmetric hybrid VLC link worths investigation. Another direction for future research is to investigate the issues related to the spatial reuse of VLC links. This requires the utilization of multiple VLC front-ends. Given the benefits and results described in this work, VLC is a promising and evolutionary wireless technology that offers valuable contribution as part of next generation heterogeneous wireless networks.

\section*{Acknowledgment}
This work was supported in part by the NSF grant ECCS-1331018 and by the Engineering Research Centers Program of the National Science Foundation under NSF Cooperative Agreement No. EEC-0812056.

\bibliographystyle{IEEEtran}
\bibliography{Hybrid_System}

\begin{thebibliography}{10}
\providecommand{\url}[1]{#1}
\csname url@samestyle\endcsname
\providecommand{\newblock}{\relax}
\providecommand{\bibinfo}[2]{#2}
\providecommand{\BIBentrySTDinterwordspacing}{\spaceskip=0pt\relax}
\providecommand{\BIBentryALTinterwordstretchfactor}{4}
\providecommand{\BIBentryALTinterwordspacing}{\spaceskip=\fontdimen2\font plus
\BIBentryALTinterwordstretchfactor\fontdimen3\font minus
  \fontdimen4\font\relax}
\providecommand{\BIBforeignlanguage}[2]{{%
\expandafter\ifx\csname l@#1\endcsname\relax
\typeout{** WARNING: IEEEtran.bst: No hyphenation pattern has been}%
\typeout{** loaded for the language `#1'. Using the pattern for}%
\typeout{** the default language instead.}%
\else
\language=\csname l@#1\endcsname
\fi
#2}}
\providecommand{\BIBdecl}{\relax}
\BIBdecl

\bibitem{kavehrad2013optical}
M.~Kavehrad, ``Optical wireless applications: A solution to ease the wireless
  airwaves spectrum crunch,'' in \emph{SPIE OPTO}.\hskip 1em plus 0.5em minus
  0.4em\relax International Society for Optics and Photonics, 2013, pp.
  86\,450G--86\,450G.

\bibitem{kahn1997wireless}
J.~M. Kahn and J.~R. Barry, ``Wireless infrared communications,''
  \emph{Proceedings of the IEEE}, vol.~85, no.~2, pp. 265--298, 1997.

\bibitem{komine2004fundamental}
T.~Komine and M.~Nakagawa, ``Fundamental analysis for visible-light
  communication system using led lights,'' \emph{Consumer Electronics, IEEE
  Transactions on}, vol.~50, no.~1, pp. 100--107, 2004.

\bibitem{rajagopal2012ieee}
S.~Rajagopal, R.~D. Roberts, and S.-K. Lim, ``{IEEE} 802.15. 7 visible light
  communication: modulation schemes and dimming support,'' \emph{Communications
  Magazine, IEEE}, vol.~50, no.~3, pp. 72--82, 2012.

\bibitem{rahaim2011hybrid}
M.~B. Rahaim, A.~M. Vegni, and T.~D. Little, ``A hybrid radio frequency and
  broadcast visible light communication system.'' in \emph{GLOBECOM Workshops},
  2011, pp. 792--796.

\bibitem{lee2013performance}
C.~Lee, C.~Tan, H.~Wong, and M.~Yahya, ``Performance evaluation of hybrid {VLC}
  using device cost and power over data throughput criteria,'' in \emph{SPIE
  Optical Engineering+ Applications}.\hskip 1em plus 0.5em minus 0.4em\relax
  International Society for Optics and Photonics, 2013, pp. 88\,451A--88\,451A.

\bibitem{chowdhury2013energy}
H.~Chowdhury, I.~Ashraf, and M.~Katz, ``Energy-efficient connectivity in hybrid
  radio-optical wireless systems,'' in \emph{Wireless Communication Systems
  (ISWCS 2013), Proceedings of the Tenth International Symposium on}.\hskip 1em
  plus 0.5em minus 0.4em\relax VDE, 2013, pp. 1--5.

\bibitem{huang2013design}
Z.~Huang and Y.~Feng, ``Design and demonstration of room division
  multiplexing-based hybrid vlc network,'' \emph{Chinese Optics Letters},
  vol.~11, no.~6, p. 060603, 2013.

\bibitem{schmid2013led}
S.~Schmid, G.~Corbellini, S.~Mangold, and T.~R. Gross, ``{LED-to-LED} visible
  light communication networks,'' in \emph{Proceedings of the fourteenth ACM
  international symposium on Mobile ad hoc networking and computing}.\hskip 1em
  plus 0.5em minus 0.4em\relax ACM, 2013, pp. 1--10.

\bibitem{langer2007recent}
K.-D. Langer and J.~Grubor, ``Recent developments in optical wireless
  communications using infrared and visible light,'' in \emph{Transparent
  Optical Networks, 2007. ICTON'07. 9th International Conference on},
  vol.~3.\hskip 1em plus 0.5em minus 0.4em\relax IEEE, 2007, pp. 146--151.

\bibitem{ramaboli2012bandwidth}
A.~L. Ramaboli, O.~E. Falowo, and A.~H. Chan, ``Bandwidth aggregation in
  heterogeneous wireless networks: A survey of current approaches and issues,''
  \emph{Journal of Network and Computer Applications}, vol.~35, no.~6, pp.
  1674--1690, 2012.

\bibitem{koudouridis2005generic}
G.~P. Koudouridis, R.~Ag{\"u}ero, E.~Alexandri, J.~Choque, K.~Dimou, H.~Karimi,
  H.~Lederer, J.~Sachs, and R.~Sigle, ``Generic link layer functionality for
  multi-radio access networks,'' in \emph{IST Mobile and Wireless
  Communications Summit}, 2005.

\bibitem{koudouridis2005switched}
G.~P. Koudouridis, H.~R. Karimi, and K.~Dimou, ``Switched multi-radio
  transmission diversity in future access networks,'' in \emph{VTC2005-fall:
  2005 {IEEE} 62nd vehicular technology conference, 1-4, proceedings}, 2005,
  pp. 235--239.

\bibitem{yaver2009performance}
A.~Yaver and G.~P. Koudouridis, ``Performance evaluation of multi-radio
  transmission diversity: {QoS} support for delay sensitive services,'' in
  \emph{Vehicular Technology Conference, 2009. VTC Spring 2009. IEEE
  69th}.\hskip 1em plus 0.5em minus 0.4em\relax IEEE, 2009, pp. 1--5.

\bibitem{kim2008mac}
J.-O. Kim, T.~Ueda, and S.~Obana, ``Mac-level measurement based traffic
  distribution over ieee 802.11 multi-radio networks,'' \emph{Consumer
  Electronics, IEEE Transactions on}, vol.~54, no.~3, pp. 1185--1191, 2008.

\bibitem{kim2010feedback}
J.-O. Kim, ``Feedback-based traffic splitting for wireless terminals with
  multi-radio devices,'' \emph{consumer electronics, IEEE Transactions on},
  vol.~56, no.~2, pp. 476--482, 2010.

\bibitem{kim2012splitting}
J.-O. Kim, P.~Davis, T.~Ueda, and S.~Obana, ``Splitting downlink multimedia
  traffic over {WiMax} and {WiFi} heterogeneous links based on
  airtime-balance,'' \emph{Wireless Communications and Mobile Computing},
  vol.~12, no.~7, pp. 598--614, 2012.

\bibitem{senie2002using}
D.~Senie, ``Using the {SOCK-PACKET} mechanism in linux to gain complete control
  of an ethernet interface,'' \emph{Internet:.(Retrieved Apr. 24, 2002)}, 2002.

\bibitem{shaoindoor}
S.~Shao, A.~Khreishah, M.~B. Rahaim, H.~Elgala, M.~Ayyash, T.~D. Little, and
  J.~Wu, ``An indoor hybrid {WiFi-VLC} internet access system,'' \emph{CARTOON
  Workshop of CellulAR Traffic Offloading to Opportunistic Networks,
  Philidelphia, PA}, October, 2014.

\bibitem{davis2011linux}
T.~Davis, W.~Tarreau, C.~Gavrilov, C.~N. Tindel, J.~Girouard, and J.~Vosburgh,
  ``Linux ethernet bonding driver howto,'' \emph{available as download
  (bonding. txt) from the Linux Channel Bonding project Web site
  http://sourceforge. net/projects/bonding/(November 2007)}, 2011.

\bibitem{bianchi1996performance}
G.~Bianchi, L.~Fratta, and M.~Oliveri, ``Performance evaluation and enhancement
  of the {CSMA/CA} mac protocol for 802.11 wireless lans,'' in \emph{Personal,
  Indoor and Mobile Radio Communications, 1996. PIMRC'96., Seventh IEEE
  International Symposium on}, vol.~2.\hskip 1em plus 0.5em minus 0.4em\relax
  IEEE, 1996, pp. 392--396.

\bibitem{langeroptoelectronics}
K.-D. Langer, J.~Hilt, D.~Shulz, F.~Lassak, F.~Hartlieb, C.~Kottke, L.~Grobe,
  V.~Jungnickel, and A.~Paraskevopoulos, ``Optoelectronics \& communications
  rate-adaptive visible light communication at {500Mb/s} arrives at plug and
  play,'' \emph{SPIE Newsroom}, 14 November 2013.

\bibitem{grobe2013high}
L.~Grobe, A.~Paraskevopoulos, J.~Hilt, D.~Schulz, F.~Lassak, F.~Hartlieb,
  C.~Kottke, V.~Jungnickel, and K.-D. Langer, ``High-speed visible light
  communication systems,'' \emph{Communications Magazine, IEEE}, vol.~51,
  no.~12, pp. 60--66, 2013.

\bibitem{tirumala2005iperf}
A.~Tirumala, F.~Qin, J.~Dugan, J.~Ferguson, and K.~Gibbs, ``Iperf: The
  {TCP/UDP} bandwidth measurement tool,'' \emph{htt p://dast. nlanr.
  net/Projects}, 2005.

\end{thebibliography}

\end{document}